\documentclass[aps,pre,twocolumn]{revtex4}
\usepackage{graphicx}

\begin{document}

\title{Equation of state for hard square lattice gases}

\author{Heitor C. Marques Fernandes} 
\email{heitor@if.ufrgs.br}
\affiliation{Instituto de F{\'\i}sica, Universidade Federal do
Rio Grande do Sul \\ CP 15051, 91501-970 Porto Alegre RS, Brazil}

\author{Yan Levin} 
\email{levin@if.ufrgs.br}
\affiliation{Instituto de F{\'\i}sica, Universidade Federal do
Rio Grande do Sul \\ CP 15051, 91501-970 Porto Alegre RS, Brazil}

\author{Jeferson J. Arenzon} 
\email{arenzon@if.ufrgs.br}
\affiliation{Instituto de F{\'\i}sica, Universidade Federal do
Rio Grande do Sul \\ CP 15051, 91501-970 Porto Alegre RS, Brazil}

\maketitle

Almost forty years ago, Carnahan and Starling published in this journal
their, now famous, equation of state for hard sphere fluid~\cite{CaSt69}. 
Their derivation
was based on the simple observation that the leading order virial 
coefficients for hard sphere fluid in three dimensions closely
followed a geometric sequence.  The assumption that this
behavior also extrapolated to
higher order virials, allowed Carnahan and Starling to
explicitly resum the virial expansion to find a simple, yet,
very accurate equation of state. 

Unfortunately, no such accurate equation of state is known for the 
case of lattice gases. This is particularly frustrating, since 
lattice
models are widely used to study many complex fluids ranging
from microemulsions to electrolytes~\cite{Wi84,KaLo99,MaDi99,ArKoKo03,Pa05}.  
In this note, we shall present
a very simple equation of state which works very well for
two dimension lattice gas of hard squares and reasonably well for
three dimension lattice gas of small hard cubes at not too high density.   

Our discussion is based on a lattice 
theory of polymer mixtures
proposed a long time ago by Flory~\cite{Flo53},
who deduced the entropy of mixing to be
%----------------------------
\begin{equation}
\label{3}
S =-k_B [N_1 \ln \phi_1 + N_2 \ln \phi_2] \;,
\end{equation}
%-------------------------------
where $N_1$ and $N_2$ are the number of polymers of type one and two,  
while  $\phi_1$  and  $\phi_2$ are their respective volume 
fractions. 
The form of Eq.(\ref{3}) is particularly appealing since it does
not contain any reference to the lattice structure and depends only
on thermodynamically well defined variables. 
The mixture is assumed to fill all the
available volume, so that there are no vacancies.
If there is only one type of polymer occupying a volume fraction $\phi_1$, 
the rest of the space is taken to be filled by the 
solvent of $\phi_2=1-\phi_1$.  

It is clear that the formalism developed by Flory for polymer
mixtures should be readily applicable to ``hard'' 
non-attracting lattice gases.  
Consider,
for example, a lattice gas of hard hypercubes of 
volume $\lambda^d$ ($\lambda$ is integer and the lattice
spacing is taken to be $1$) 
on a simple hypercubic lattice in $d$ dimensions.
The Helmholtz free energy of this lattice gas is
$F_m=-T S$, since the system is athermal.
The free energy density is
%-----------------------------
\begin{equation}
\label{4}
\beta f_m = \rho \ln \phi+\left( 1 - \phi \right) \ln \left( 1 - \phi \right) \;,  
\end{equation}
%-----------------------------------------
where $\rho$ is the particle density 
and $\phi = \lambda^d \rho$ is the 
volume fraction.  

We note, however, 
that in the low density limit,  Eq. (\ref{4}) {\it does not} 
reduce to the free energy of the ideal gas 
%-----------------------------
\begin{equation}
\label{5}
\beta f = \rho \ln \rho -\rho \;.  
\end{equation}
%-----------------------------------------
Therefore,  $f_m$  can not be the {\it total}
free energy of the system, except for the case of  $\lambda=1$ when 
Eq. (\ref{4}) becomes exact.  
For polymer mixtures, 
to obtain the total free energy, Flory added an
extra contribution to Eq.(\ref{4}) which accounted  
for the conformational degrees of
freedom of the polymer chains, the so called entropy of 
disorientation~\cite{Flo53}.
This restored the correct low density behavior to the
theory.  For rigid particles, however, 
the entropy of disorientation
is identically zero and 
cannot be the reason for the failure of Eq. (\ref{4}). 

To recover the correct low density behavior, 
while preserving the simple and thermodynamically appealing
form of Eq. (\ref{4}), we  modify $f_m$ 
by introducing a multiplicative factor $g(\lambda)$ into Eq. (\ref{4}),
%------------------------------------- 
\begin{equation}
\label{6}
\beta f = \rho \ln \phi+g\left( \lambda \right)
\left( 1 - \phi \right) \ln \left( 1 - \phi \right). 
\end{equation}
%-----------------------------------
This equation can be interpreted as an interpolation between
the low density limit governed by the particles, and the high
density limit in which defects, the ``holes'', become relevant.
The total number of holes, however, 
is not fixed since the vacancies can change their size and shape,
so that the prefactor appearing
in front of the second term of Eq. (\ref{6}) is 
the effective number density.

The requirement that in the low density limit  
Eq. (\ref{6}) must reduce to  Eq. (\ref{5}), uniquely determines the
functional form of $g(\lambda)$ yielding
%-------------------------
\begin{equation}
\label{7}
g\left( \lambda \right) = \frac{1+d \ln \lambda}{\lambda^d} \;.
\label{eq.glambda}
\end{equation}
%---------------------------
Note that $g(1)=1$, so that Eq.~(\ref{6}) reduces to the exact free energy
for a lattice gas of $\lambda=1$.
The chemical potential within the modified Flory approximation (MFA) is
%--------------------------------
\begin{equation}
\label{8}
\beta \mu = - (1+d \ln \lambda) \left[ \ln \left( 1 - \phi\right)
+ 1  \right]+ \ln \phi + 1 \;.
\end{equation}
%-----------------------------------

\begin{figure}[floatfix]
\includegraphics[width=7.5cm]{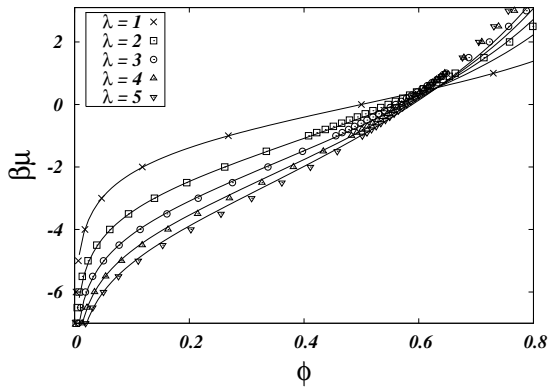}
\caption{Chemical potential versus volume fraction for various
two dimensional lattice gases. The symbols are the MC results while the 
lines are the predictions of the MFA Eq. (\ref{8}), with $d=2$.}
\label{fig1}
\end{figure}

\begin{figure}[floatfix]
\includegraphics[width=7.5cm]{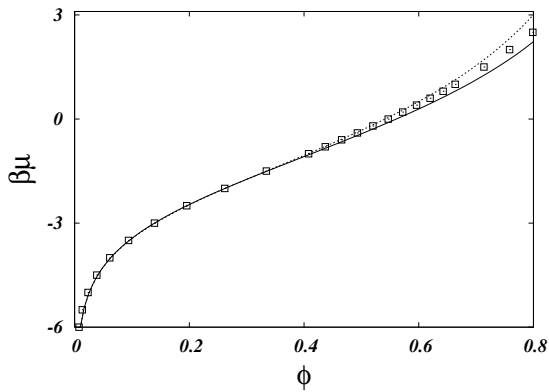}
\caption{Chemical potential versus volume fraction for $\lambda=2$
lattice gas. The symbols are the MC results; the solid curve is the
equation of state obtained in this Note, Eq.~(\ref{8});  the dashed
line is the equation of state obtained using the Fundamental Measure Density
Functional Theory, Ref.~\cite{LaCu03}, which is also the same as the one
found earlier by Temperley~\cite{Te61}. The order-disorder transition occurs
at $\phi \approx 0.93$.}
\label{fig1a}
\end{figure}

In Fig.~\ref{fig1} we compare the value of the chemical 
potential obtained
within the MFA
with the results of the Monte Carlo simulations for a gas
of hard squares of different sizes $\lambda$.
The simulations were performed using the grand-canonical
ensemble at fixed volume $V$, temperature $T$, 
and chemical potential $\mu$, with  
trial moves insertion and removal of 
particles as well as attempts to diffuse~\cite{FrSm02}.
%--------------------------
%--------------------------------
The agreement is excellent for all $\lambda$s tested, up
to quite high volume fractions.  However, similarly to the 
Carnahan-Starling equation
of state,  MFA also fails to notice the phase transition between
the disordered and ordered (columnar) phases present at high volume
fractions~\cite{Run72}.  In Fig. \ref{fig1a} we compare the accuracy of the 
MFA with the earlier equation of state derived by 
Temperley~\cite{Te61} which is also identical to the one recently found
using the Fundamental Measure theory~\cite{LaCu03}. 

It is curious that  all the 
MC curves for different values of $\lambda$  intersect
at approximately the same point.  
This property is also captured by the MFA, which
predicts that all the chemical potentials for different  
$\lambda$s are equal when the volume fraction satisfies
$\ln(1-\phi_\times)= -1$, independent of $d$.
The value $\phi_\times=0.632121$ is in excellent agreement 
with the intersection point observed in the Monte Carlo 
simulations.
%--------------------------
%--------------------------------

\begin{figure}
\includegraphics[width=7.5cm]{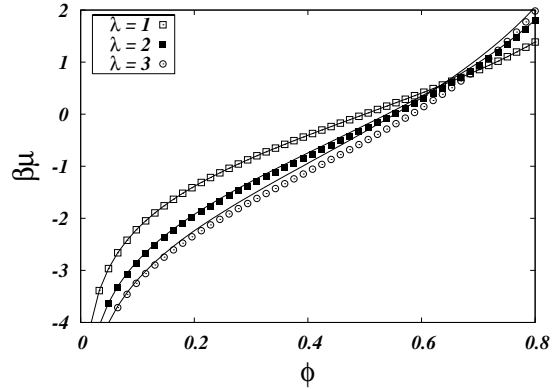}
\caption{Chemical potential versus volume fraction for a gas of 
hard rods with $\lambda=1,2$ and $3$ in $d=1$. 
The symbols are the exact value of the
chemical potential~\cite{Ro80} while the 
lines are the predictions of the MFA Eq. (\ref{8}) with $d=1$.}
\label{fig2}
\end{figure}

\begin{figure}
\includegraphics[width=7.5cm]{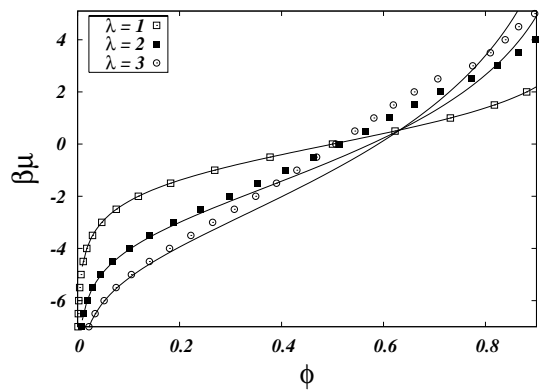}
\caption{Chemical potential versus volume fraction for various
three dimensional hard cube lattice gases. The points 
are the MC results while the 
lines are the predictions of the MFA Eq. (\ref{8}) with $d=3$.}
\label{fig3}
\end{figure}

In Figs. \ref{fig2} and \ref{fig3} we also show the equations of state for $d=1$ hard rods,
and $d=3$ hard cubes.  
In the case of $d=1$ the exact free energy is known~\cite{Ro80}.
Although still quite good, the agreement between the simulations and the MFA
deteriorates more rapidly with increasing $\lambda$ 
for $d=1$ and $d=3$ than for $d=2$. 
%--------------------------
%--------------------------------

%--------------------------
%--------------------------------
The high degree of accuracy of the MFA in $d=2$ is quite surprising
in view of the crudeness of the approximation.  
It also suggests
that there should be a more direct way to arrive at 
Eq.~(\ref{6}), or some other such 
equation of state~\cite{LaCu03,LaCu04}, generally
valid for non-attracting lattice gases of arbitrary $\lambda$.  
In the absence of such general theory, the very simple Eq.~(\ref{6}) 
should be useful for constructing lattice 
mean-field theories for various
complex systems.
  
This work was supported in part by the Brazilian science agencies
CNPq, CAPES, and FAPERGS.

%%%%%%%%%%%%%%%%%%%%%%%%%%%%%%%%%%%%%%%%%%%%%%%%%%%%%%%%%%%%%%%%%%%%%%%
%                   bibliografia                                      %
%%%%%%%%%%%%%%%%%%%%%%%%%%%%%%%%%%%%%%%%%%%%%%%%%%%%%%%%%%%%%%%%%%%%%%%
\bibliographystyle{prsty}
\bibliography{references}

\begin{thebibliography}{10}

\bibitem{CaSt69}
N.~F. Carnahan and K.~E. Starling, J. Chem. Phys. {\bf 51},  635  (1969).

\bibitem{Wi84}
B. Widom, J. Chem. Phys. {\bf 81},  1030  (1984).

\bibitem{KaLo99}
E. Caglioti and V. Loreto, Phys. Rev. Lett. {\bf 83},  4333  (1999).

\bibitem{ArKoKo03}
M.~N. Artyomov, V. Kobelev, and A.~B. Kolomeisky, J. Chem. Phys. {\bf 118},
  6394  (2003).

\bibitem{Pa05}
A.~Z. Panagiotopoulos, J. Chem. Phys. {\bf 123},  Art. No. 104504  (2005).

\bibitem{MaDi99}
J. Marro and R. Dickman, {\em Nonequilibrium phase transitions in lattice
  models} (Cambridge University Press, ADDRESS, 1999).

\bibitem{Flo53}
P.~J. Flory, {\em Principles of polymer chemistry} (Cornell University Press,
  Ithaca, New York, 1953).

\bibitem{FrSm02}
D. Frenkel and B. Smit, {\em Understanding {M}olecular {S}imulation} (Academic
  Press, New York, 2002).

\bibitem{Run72}
L. Runnels,  in {\em Phase transitions and critical phenomena.}, edited by C.
  Domb and M.~S. Green (Academic Press, London-New York, 1972), Vol.~2,
  Chap.~8, pp.\ 305--328, lattice gas theories of melting.

\bibitem{Te61}
H. Temperley, P. Phys. R. Soc. {\bf 77},  630  (1961).

\bibitem{LaCu03}
L. {Lafuente} and J. {Cuesta}, J. Chem. Phys. {\bf 119},  10832  (2003).

\bibitem{Ro80}
A. Robledo, J. Chem. Phys. {\bf 72},  1701  (1980).

\bibitem{LaCu04}
L. Lafuente and J.~A. Cuesta, Phys. Rev. Lett. {\bf 93},  130603  (2004).

\end{thebibliography}

\end{document}